\documentclass[11pt]{article}
\usepackage{my_aaspp4}
\usepackage{amsmath}
\usepackage{graphics}

\DeclareRobustCommand{\ion}[2]{%
\relax\ifmmode
\ifx\testbx\f@series
{\mathbf{#1\,\mathsc{#2}}}\else
{\mathrm{#1\,\mathsc{#2}}}\fi
\else\textup{#1\,{\mdseries\textsc{#2}}}%
\fi}

\def\degr{\hbox{$^\circ$}}

\def\nii{[\ion{N}{ii}]}
\begin{document}

%
   \title{ON THE NATURE OF THE FBS BLUE STELLAR OBJECTS AND THE COMPLETENESS 
   OF THE BRIGHT QUASAR SURVEY. II.}

	\author{A.\,M. Mickaelian}	
   	\affil{Byurakan Astronomical Observatory, Armenian national Academy of 
Sciences, \\
378433 Byurakan, Armenia}
	\author{A.\,C. Gon\c{c}alves}
	\affil{ESO, Karl Schwarzschild Strasse 2, D-85748 Garching bei M\"unchen, 
	 Germany}
	\author{M.-P. V\'eron-Cetty and P. V\'eron}
	\affil{Observatoire de Haute Provence, CNRS, F-04870 Saint-Michel 
	l'Observatoire, France}






\vspace{1.5cm}

   \begin{abstract}
 In Paper I (Mickaelian et al. 1999), we compared the surface density of QSOs 
in the Bright Quasar Survey (BQS) and in the First Byurakan Survey (FBS) and 
concluded that the completeness of the BQS is of the order of 70\/\% rather than 
30--50\/\% as suggested by several authors. A number of new observations recently 
became available, allowing a re-evaluation of this completeness. We now obtain a
surface density of QSOs brighter than $B =$ 16.16 in a subarea of the FBS covering
$\sim$2\,250 deg$^{2}$, equal to 0.012 deg$^{-2}$ (26~QSOs), implying a 
completeness of 53$\pm$10\/\%.

   \end{abstract}

      \keywords{ Quasars -- Surveys }

%
\vspace{1cm}
\section{INTRODUCTION}

 In Paper I, by comparing the FBS (Markarian et al. 1989) and BQS (Green et al.
1986) surveys in their area in common, we derived a completeness of 
$\sim$70\/\% for the BQS. A 
number of bright AGNs have since been discovered in the area which, together 
with our new spectroscopic observations, allowed us to refine our previous 
estimate of the BQS completeness.

\section{OBSERVATIONS}

 We have obtained new spectra for 11 FBS objects. The observations were carried
out on November 25, 1998 and January 14--15, 1999 at the Byurakan Astrophysical 
Observatory (BAO) and at the Observatoire de Haute-Provence (OHP), respectively.
The journal of observations is given in Table 1, together with relevant data. \\

\begin{table}[t]
\caption{\label{spectra}New spectra. Col. 1 gives the name, col. 2 the FBS 
number, col. 3 the original FBS classification, col. 4 the magnitude, cols. 5 
and 6 the place and date of observation, col. 7  the galactic latitude, col. 8 
the classification and col. 9 the redshift} 
\begin{center}
\begin{tabular}{lrllllrll}
\hline
& FBS \# & & mag & & ~~date & $b$~~ & & ~~$z$ \\
\hline
 FBS\,0228+447 &    227~~ & B2  & 15.5 & BAO & 25.11.98 & $-$14.4 & *  &       \\
 FBS\,0744+818 & 1\,055~~ & B1e & 16.4 & OHP & 14.01.99 &    29.1 & *  &       \\
 FBS\,0747+729 &    966~~ & N3e & 15.8 & BAO & 25.11.98 &    30.4 & *  &       \\
 FBS\,0929+733 &    876~~ & B3e & 16.3 & BAO & 25.11.98 &    37.2 & *  &       \\
 FBS\,0944+713 &    878~~ & B3e & 18.6 & OHP & 14.01.99 &    39.3 & *  &       \\
 FBS\,0950+664 &    785~~ & N2  & 16.7 & BAO & 25.11.98 &    42.4 & S1 & 0.172 \\
 FBS\,1049+803 & 1\,068~~ & N1e & 17.1 & BAO & 25.11.98 &    35.7 & *  &       \\
 FBS\,1235+699 &    894~~ & N1e & 17.9 & OHP & 15.01.99 &    47.4 & Q  & 0.521 \\
 FBS\,1324+448 &    322~~ & B1  & 17.  & OHP & 15.01.99 &    71.1 & Q  & 0.331 \\
 FBS\,1715+406 &    936~~ & se  & 16.  & OHP & 18.01.99 &    34.5 & G  & 0.029 \\
 FBS\,2308+425 &    418~~ & B1  & 13.5 & OHP & 14.01.99 & $-$16.3 & *  &       \\

\hline
\end{tabular}
\end{center}
\end{table}
\normalsize

  Seven of the newly observed objects are stars, including FBS\,2308$+$425 (at 
$b = -$16.3\degr) which is associated with a ROSAT RASS (Voges et al. 1999) 
X-ray source (Table 2, Paper I). 

 FBS\,0950$+$664 (RXS J09540$+$6608) has been identified on an objective prism
plate as an AGN (Bade et al. 1998); our spectrum shows it to be a Seyfert 1 at 
$z =$ 0.172. Our new spectra of FBS\,1235$+$699 and FBS\,1324$+$448 confirm 
their redshift ($z =$ 0.521 and 0.331 respectively). FBS\,1715$+$406 is 
Zw\,225.094 or MCG\,07.35.061, a 15.4 mag galaxy at $z =$ 0.030 (Marzke et al. 
1996); according to Abramian \& Mickaelian (1994), it is an emission line 
galaxy; our spectrum shows that it is an absorption line galaxy, with a weak 
\nii\,$\lambda$6583 line in emission at $z =$ 0.029. 

 The spectra of the four extragalactic objects are displayed in 
Fig.~\ref{spectra}.\\

\begin{figure}[t]
\begin{center}
\resizebox{14cm}{!}{\includegraphics{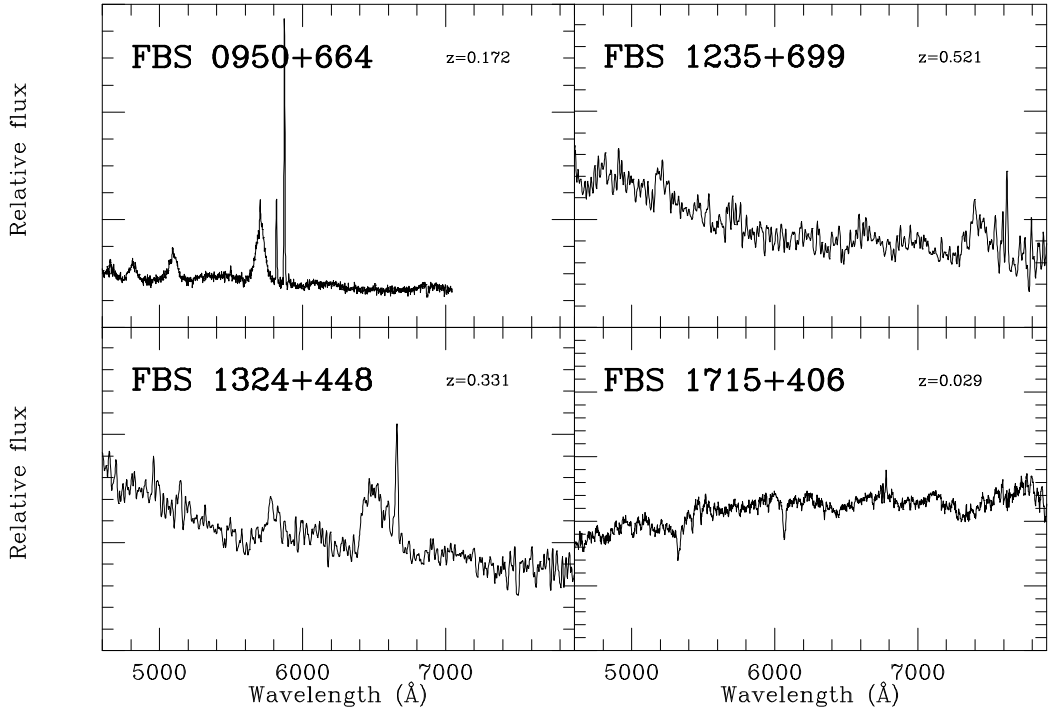}}
\caption{\label{spectra}
Spectra of the four extragalactic objects in Table \ref{spectra}.
}
\end{center}
\end{figure}

\section{NEW PUBLISHED DATA}

 Since the publication of Paper I\footnotemark\, 51 new bright ($B <$ 17.0) 
AGNs have been discovered, at $|b| >$ 30\degr, in the subarea of the FBS survey studied
in this paper, bringing the total number of known such objects to 108.

\begin{figure}[t]
\begin{center}
\resizebox{9cm}{!}{\includegraphics{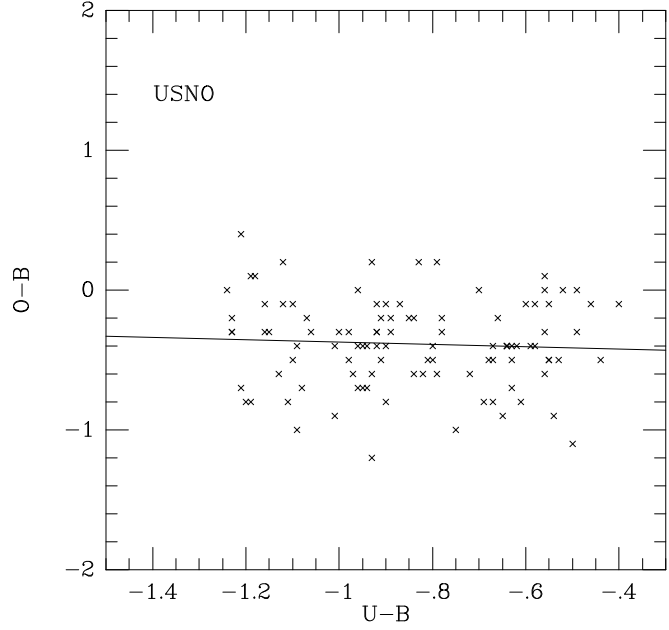}}
\caption{\label{usno_mag}Plot of the differences between the USNO $O$ and the 
photoelectric $B$ magnitudes {\it vs} the photoelectric $U - B$ colors for 102 PG
objects}
\end{center}
\end{figure}

\begin{figure}[h!]
\begin{center}
\resizebox{9cm}{!}{\includegraphics{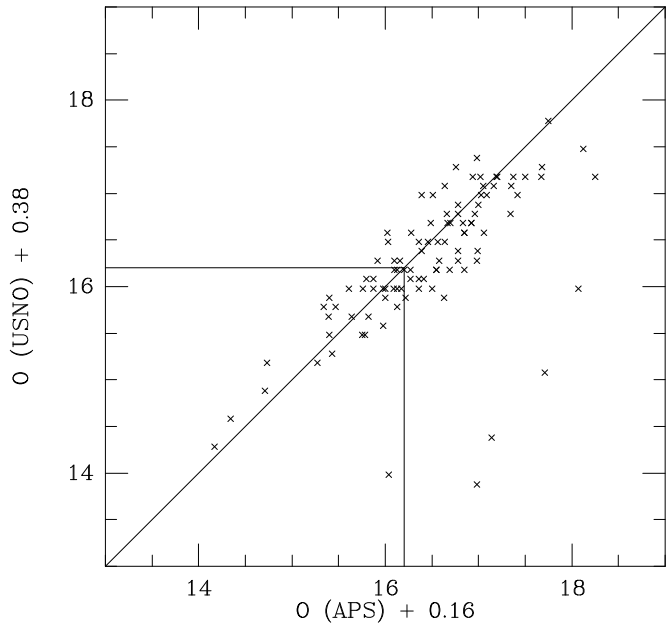}}
\caption{\label{usno_aps}Plot of the USNO {\it vs} the APS $O$ magnitudes for 
the bright AGNs listed in tables 2 and 3}
\end{center}
\end{figure}

\footnotetext[1]{In Paper I, we claimed that the position accuracy of the 
FBS objects in the last seven papers by Abramian \& Mickaelian 
is much better than in the first four papers of the series; in fact, the 
objects \#924 to \#939 in paper IX (Abramian \& Mickaelian 1994) have an
accuracy as poor as in the first four papers.}

\subsection{MAGNITUDE ESTIMATE}
 We have extracted, when available, the $O$ magnitudes of these 108 objects from 
the APS database (Pennington et al. 1993); these magnitudes are missing for five
objects only (all at $\delta >$~63\degr). We have also extracted the $O$
magnitudes from the USNO-A2.0 catalogue (Monet et al. 1996) which, as the APS, is 
based on measurements of the $O$ plates of the Palomar Observatory Sky Survey I 
(POSS-I). 

 To estimate the accuracy of these magnitudes, we have proceeded as for the APS 
$O$ magnitudes (see Paper I): we have compared the differences between the
USNO $O$ and photoelectric $B$ magnitudes of 102 PG UV-excess stars {\it vs} their 
photoelectric $U - B$ colours (Fig.~\ref{usno_mag}); we found a negligible colour 
equation, a {\it rms} dispersion of 0.31 mag, compared to 0.25 mag for the APS 
magnitudes, and a relatively large offset $<$$O - B$$>$ $= -$0.38 mag (to be 
compared with $<$$O - B$$>$~$=~-$0.16 mag for the APS). Fig.~\ref{usno_aps} shows 
a comparison of the APS and USNO $O$ magnitudes for the bright AGNs; they are in 
reasonable agreement, except for five objects for which the USNO magnitudes are 
brighter by more than one mag than the APS magnitudes; these five objects are 
low luminosity Seyfert 1 galaxies at relatively small redshifts ($z <$ 0.13) 
which probably explain the large magnitude differences; it seems that the USNO 
magnitudes for extended objects are grossly underestimated. \\

 In Table \ref{brightagn}, we list all bright ($B <$ 17.0) AGNs found in the FBS
subarea at $|b| >$ 30\degr\ with their APS and USNO $O$ magnitudes (when 
available) and the absolute $B$ magnitudes computed\footnotemark[2]\ using the 
APS $O$ magnitudes increased by 0.16 mag (or the USNO $O$ magnitudes increased by 
0.4 mag), excluding the bright QSOs of our ``complete" sample (listed in 
Table~\ref{brightqso}). 

\footnotetext[2]{assuming $H_{\rm o} =$ 50\,km\,s$^{-1}$\,Mpc$^{-1}$}

\begin{table}
\caption{\label{brightagn}Bright AGNs ($B <$ 17.0) in the FBS subarea at 
$|b| >$ 30\degr, excluding the bright QSOs listed in Table 3. Cols. 1 to
6 give the B1950 position, cols. 7 and 8, the APS and USNO-A2.0 $O$ magnitudes 
respectively, col. 9 the name, col. 10 the redshift, col. 11 the galactic 
latitude, col. 12 the absolute $B$ magnitude and col. 13 references for the newly 
identified AGNs:
(1) Beuermann et al. (1999), (2) Cao et al. (1999), (3) Wei et al. (1999),  
(4) White et al. (2000), (5) Schwope et al. (2000), (6) Xu et al. (1999), 
(7) present paper
} 
\small
\begin{center}
\begin{tabular}{lrrrrrrccllccc}
\hline
\multicolumn{4}{c}{~~$\alpha$(B1950)} & 
\multicolumn{3}{c}{$\delta$(B1950)} & APS $O$ & US $O$ & Name & ~~~$z$ & $b$ & ~$M_{\rm B}$ & ref.   \\
\hline
     &  8 &  6 &34.3 & 62 & 12 & 11 &16.11 &15.8 & HS 0806+6212    & 0.173 &  33.0 & $-$23.8 &     \\
     &  8 & 38 &31.7 & 77 &  3 & 59 &17.55 &14.7 & FBS 0838+771    & 0.131 &  32.7 & $-$21.8 &     \\
     &  8 & 38 &46.5 & 40 & 29 & 16 &16.35 &16.6 & RXS J08420+4018 & 0.151 &  37.6 & $-$23.3 & (5) \\
     &  8 & 44 &33.9 & 34 & 56 &  9 &16.98 &14.0 & FBS 0844+349    & 0.064 &  38.0 & $-$20.7 &     \\
     &  8 & 51 &28.5 & 40 & 30 & 34 &16.23 &16.6 & RXS J08547+4019 & 0.152 &  40.0 & $-$23.5 & (3) \\
     &  8 & 53 &58.4 & 75 &  6 & 34 &16.40 &16.1 & RXS J08595+7455 & 0.276 &  34.3 & $-$24.6 & (2) \\
     &  9 & 24 &39.6 & 40 & 51 & 37 &16.25 &16.0 & KUV 09247+4052  & 0.419 &  46.3 & $-$25.6 & (6) \\
     &  9 & 31 &50.8 & 43 & 44 & 35 &16.47 &15.5 & FBS 0931+437    & 0.456 &  47.4 & $-$25.6 &     \\
     &  9 & 35 &48.7 & 41 & 41 & 55 &16.06 &15.5 & FBS 0935+416    & 1.966 &  48.3 & $-$29.7 &     \\
     &  9 & 36 &38.9 & 39 & 37 & 38 &16.69 &16.2 & PG 0936+396     & 0.458 &  48.6 & $-$25.4 &     \\
     &  9 & 45 &33.2 & 40 & 44 & 42 &16.82 &13.5 & NPM1G+40.0197   & 0.047 &  50.2 & $-$20.3 & (5) \\
     &  9 & 47 &44.8 & 39 & 40 & 54 &16.39 &15.8 & FBS 0947+396    & 0.206 &  50.7 & $-$23.9 &     \\
     &  9 & 50 & 9.5 & 66 & 22 & 31 &17.00 &16.7 & FBS 0950+664    & 0.172 &  42.4 & $-$22.9 & (7) \\
     &  9 & 59 & 9.6 & 68 & 27 & 48 &16.01 &15.6 & FBS 0959+685    & 0.773 &  42.0 & $-$27.3 &     \\
     & 10 &  2 &12.0 & 34 & 28 & 59 &16.86 &16.8 & FIRST J1005+3414& 0.162 &  53.8 & $-$23.0 & (4) \\
     & 10 &  2 &37.3 & 43 & 47 & 17 &16.39 &15.8 & FBS 1002+437    & 0.178 &  52.9 & $-$23.6 &     \\
     & 10 & 16 & 4.7 & 34 & 51 & 35 &16.62 &16.5 & FIRST J1018+3436& 0.109 &  56.6 & $-$22.3 & (4) \\
     & 10 & 22 &53.7 & 68 &  1 & 29 &  --  &16.8 & RX J10265+6746  & 1.178 &  43.9 & $-$27.4 &     \\
     & 10 & 25 &11.2 & 63 & 18 &  8 &  --  &15.8 & RXS J10286+6302 & 0.080 &  47.2 & $-$22.2 & (5) \\
     & 10 & 28 &44.8 & 70 & 43 &  0 &  --  &16.3 & RXS J10325+7027 & 0.063 &  42.4 & $-$21.2 & (5) \\
     & 10 & 31 & 7.6 & 36 & 10 & 39 &16.76 &16.3 & CSO 275         & 0.169 &  59.6 & $-$22.6 & (4) \\
     & 10 & 32 &58.3 & 38 & 12 & 14 &16.51 &16.3 & B3 1032+382     & 1.508 &  59.6 & $-$28.5 & (4) \\
     & 10 & 48 &56.5 & 34 & 15 & 22 &15.94 &15.8 & FBS 1048+343    & 0.167 &  63.4 & $-$23.9 &     \\
     & 10 & 49 &22.4 & 61 & 41 & 18 &16.62 &16.4 & FBS 1049+617    & 0.421 &  50.4 & $-$25.2 &     \\
     & 10 & 50 &37.5 & 66 & 27 & 59 &  --  &15.9 & RXS J10539+6612 & 0.117 &  46.9 & $-$23.0 & (1) \\
     & 11 & 12 &19.3 & 66 & 48 & 23 &16.53 &15.8 & FBS 1112+668    & 0.544 &  47.9 & $-$25.9 &     \\
     & 11 & 12 &19.8 & 43 &  6 & 11 &17.03 &16.8 & PG 1112+431     & 0.302 &  64.9 & $-$24.2 &     \\
     & 11 & 17 &16.8 & 39 & 44 & 32 &15.18 &15.4 & CG 1410         & 0.086 &  67.4 & $-$23.2 & (5) \\
     & 11 & 19 &27.3 & 42 & 53 & 36 &16.80 &16.4 & CSO 1169        & 0.813 &  66.2 & $-$26.9 & (4) \\
     & 11 & 27 &15.9 & 37 &  5 & 51 &17.59 &17.4 & FIRST J1129+3649& 0.399 &  70.2 & $-$24.3 & (4) \\
\hline
\end{tabular}
\end{center}
\normalsize
\end{table}
\addtocounter{table}{-1}
\begin{table}
\caption{(continues)}
\small
\begin{center}
\begin{tabular}{lrrrrrrccllccc}
\hline
\multicolumn{4}{c}{~~$\alpha$(B1950)} & 
\multicolumn{3}{c}{$\delta$(B1950)} & APS $O$ & US $O$ & Name & ~~~$z$ & $b$ & ~$M_{\rm B}$ & ref.   \\
\hline
     & 11 & 27 &23.0 & 41 & 32 & 52 &16.20 &15.7 & KUV 11274+4133  & 1.530 &  68.1 & $-$28.4 &     \\
     & 11 & 33 &57.2 & 39 & 16 & 41 &16.11 &15.7 & FIRST J1136+3900& 0.795 &  70.4 & $-$27.5 & (4) \\
     & 11 & 34 &17.3 & 34 & 49 & 12 &17.04 &16.8 & FIRST J1136+3432& 0.192 &  72.4 & $-$23.1 & (4) \\
     & 11 & 37 & 9.3 & 66 &  4 & 28 &16.25 &15.7 & FBS 1137+661    & 0.652 &  49.7 & $-$26.6 &     \\
     & 11 & 40 &56.8 & 68 &  1 & 34 &16.82 &15.9 & FBS 1140+680    & 0.796 &  48.1 & $-$26.6 &     \\
     & 11 & 47 &46.0 & 67 & 15 & 28 &16.69 &16.2 & FBS 1147+673    & 1.020 &  49.1 & $-$27.4 &     \\
     & 11 & 48 &41.7 & 38 & 39 &  2 &16.20 &15.6 & FIRST J1151+3822& 0.336 &  73.1 & $-$25.2 & (4) \\
     & 11 & 48 &53.3 & 38 & 42 & 33 &17.34 &16.8 & B2 1148+38      & 1.304 &  73.1 & $-$27.4 & (4) \\
     & 11 & 50 &16.5 & 33 & 24 &  0 &16.30 &16.1 & FBS 1150+334    & 1.389 &  76.0 & $-$28.6 &     \\
     & 11 & 58 &17.6 & 35 & 25 & 13 &16.76 &16.3 & HS 1158+3525    & 1.700 &  76.6 & $-$28.5 & (4) \\
     & 12 &  1 &51.1 & 43 & 47 & 38 &16.23 &16.0 & FBS 1201+437    & 0.663 &  71.1 & $-$26.7 & (6) \\
     & 12 &  8 &37.8 & 70 & 22 & 12 &  --  &14.3 & RXS J12110+7005 & 0.127 &  46.6 & $-$24.8 & (5) \\
     & 12 & 11 &32.8 & 33 & 26 & 26 &17.26 &16.6 & B2 1211+33      & 1.598 &  79.9 & $-$27.9 & (4) \\
     & 12 & 18 & 5.9 & 39 &  9 & 55 &16.67 &16.3 & FIRST J1220+3853& 0.376 &  76.6 & $-$25.1 & (4) \\
     & 12 & 35 &12.9 & 69 & 58 & 13 &17.96 &17.1 & FBS 1235+699    & 0.522 &  47.4 & $-$24.4 &     \\
     & 12 & 42 &46.1 & 34 & 12 & 33 &17.52 &16.9 & FBS 1242+342    & 0.717 &  83.1 & $-$25.6 &     \\
     & 12 & 48 &26.6 & 40 &  7 & 58 &16.33 &16.3 & PG 1248+401     & 1.032 &  77.3 & $-$27.8 &     \\
     & 12 & 55 & 1.7 & 44 & 45 & 47 &16.48 &16.1 & FBS 1255+447    & 0.30  &  72.6 & $-$24.7 &     \\
     & 12 & 57 &26.8 & 34 & 39 & 31 &17.21 &16.8 & B 201           & 1.375 &  82.5 & $-$27.6 & (4) \\
     & 13 & 12 &37.0 & 42 & 34 &  9 &15.31 &15.4 & NPM1G+42.0343   & 0.073 &  74.1 & $-$22.7 & (5) \\
     & 13 & 24 &54.6 & 44 & 50 & 36 &18.09 &16.8 & FBS 1324+448    & 0.331 &  71.1 & $-$23.3 &     \\
     & 13 & 28 &40.2 & 41 & 44 & 22 &16.60 &16.9 & RXS J13308+4128 & 0.182 &  73.5 & $-$23.5 & (5) \\
     & 13 & 29 &29.8 & 41 & 17 & 23 &16.78 &16.8 & FBS 1329+412    & 1.937 &  73.8 & $-$28.9 &     \\
     & 13 & 38 &28.6 & 40 & 51 & 48 &16.82 &17.0 & RXS J13406+4036 & 0.161 &  73.1 & $-$23.0 & (5) \\
     & 13 & 38 &52.0 & 41 & 38 & 22 &16.50 &16.4 & FBS 1338+416    & 1.204 &  72.5 & $-$28.0 &     \\
     & 13 & 39 &47.8 & 37 & 22 & 16 &16.89 &16.7 & CSO 1010        & 1.106 &  75.4 & $-$27.4 & (4) \\
     & 13 & 51 &46.3 & 64 &  0 & 29 &14.55 &14.5 & FBS 1351+640    & 0.088 &  52.0 & $-$23.9 &     \\
     & 13 & 54 & 2.3 & 41 & 50 & 53 &16.69 &15.8 & RXS J13561+4136 & 0.697 &  70.4 & $-$26.4 & (4) \\
     & 14 &  0 &50.9 & 33 & 34 & 26 &16.12 &16.2 & RXS J14030+3320 & 0.342 &  73.4 & $-$25.4 & (5) \\
     & 14 & 15 &57.2 & 43 & 25 & 43 &17.51 &16.8 & RXS J14179+4311 & 0.079 &  66.2 & $-$20.7 & (5) \\
     & 14 & 16 &43.3 & 42 & 47 & 29 &16.34 &15.6 & HS 1416+4247    & 0.421 &  66.5 & $-$25.6 & (4) \\
     & 14 & 22 &57.6 & 42 & 27 & 36 &16.42 &15.9 & RX J14249+422   & 0.316 &  65.7 & $-$24.9 &     \\
     & 14 & 24 &29.2 & 39 & 17 & 10 &17.91 &15.6 & RXS J14265+3903 & 0.081 &  66.9 & $-$20.4 & (5) \\
     & 14 & 29 &20.9 & 40 &  5 & 55 &16.62 &15.9 & CSO  464        & 1.217 &  65.7 & $-$28.0 & (4) \\
     & 14 & 29 &52.1 & 34 & 30 &  2 &16.84 &16.5 & FIRST J1431+3416& 0.704 &  67.3 & $-$26.3 & (4) \\
\hline
\end{tabular}
\end{center}
\normalsize
\end{table}
\addtocounter{table}{-1}
\begin{table}[ht!]
\caption{(end)}
\begin{center}
\small
\begin{tabular}{lrrrrrrccllccc}
\hline
\multicolumn{4}{c}{~~$\alpha$(B1950)} & 
\multicolumn{3}{c}{$\delta$(B1950)} & APS $O$ & US $O$ & Name & ~~~$z$ & $b$ & ~$M_{\rm B}$ & ref.   \\
\hline
     & 15 & 21 &59.0 & 39 & 24 & 39 &16.93 &16.6 & HS 1521+3924    & 0.657 &  56.2 & $-$26.0 & (4) \\
     & 15 & 26 &52.0 & 65 & 58 & 32 &16.90 &16.2 & FBS 1526+659    & 0.345 &  44.4 & $-$24.6 &     \\
     & 15 & 43 &15.9 & 35 &  2 &  6 &17.18 &16.4 & RXS J15451+3452 & 0.518 &  52.3 & $-$25.1 & (4) \\
     & 16 & 11 &13.3 & 37 & 24 & 49 &15.88 &13.6 & MCG 06.36.003   & 0.070 &  46.7 & $-$22.0 & (5) \\
     & 16 & 12 &59.6 & 37 & 53 & 34 &16.87 &16.6 & FIRST J1614+3746& 1.532 &  46.4 & $-$28.2 & (4) \\
     & 16 & 30 &15.1 & 37 & 44 &  8 &16.62 &16.0 & FBS 1630+377    & 1.478 &  42.9 & $-$28.4 &     \\
     & 16 & 31 &19.4 & 39 & 30 & 42 &16.48 &16.7 & KUV 16313+3931  & 1.023 &  42.8 & $-$27.6 & (4) \\
     & 16 & 39 &36.8 & 35 & 56 &  0 &16.54 &16.3 & FIRST J1641+3550& 1.438 &  40.9 & $-$28.4 & (4) \\
     & 17 &  1 &36.2 & 37 & 41 & 32 &15.60 &15.6 & RXS J17033+3737 & 0.065 &  36.8 & $-$22.1 & (5) \\
     & 17 &  3 & 3.4 & 38 &  6 &  9 &16.83 &16.0 & FIRST J1704+3802& 0.063 &  36.6 & $-$20.9 & (4) \\
     & 17 &  6 &17.5 & 69 &  1 & 29 &16.04 &15.8 & HS 1706+6901    & 0.449 &  34.6 & $-$26.0 &     \\
     & 17 & 11 &17.2 & 35 & 27 &  1 &16.84 &16.3 & FIRST J1713+3253& 0.083 &  34.5 & $-$21.5 & (4) \\
     & 17 & 27 &18.3 & 38 & 40 & 46 &17.19 &16.7 & B3 1727+386     & 1.386 &  32.0 & $-$27.7 & (4) \\   
     & 17 & 32 &26.6 & 40 & 39 & 50 &16.20 &16.1 & FIRST J1734+4037& 0.356 &  31.4 & $-$25.4 & (4) \\
\hline
\end{tabular}
\end{center}
\end{table}
\normalsize

\begin{table*}
\caption{\label{brightqso}Bright QSOs ($B <$ 16, and $M_{B} < -$24) in the FBS 
subarea at $|b| >$ 30\degr.
The columns are the same as in Table 2 with however two additional columns;
an X in col. 14 indicates that the object is a ROSAT RASS source; a Y or an N
in col. 15 indicates if the object lies or not in the PG area} 
\begin{center}
\small
\begin{tabular}{rrrrrrccllcclll}
\hline
\multicolumn{3}{c}{$\alpha$(B1950)} & 
\multicolumn{3}{c}{$\delta$(B1950)} & APS $O$ & US $O$ & Name & ~~~$z$ & $b$ 
 & ~$M_{B}$ & ref. & &  \\
\hline
   8 &  4 &35.4 & 76 & 11 & 33 &14.18 &14.2 & FBS 0804+762    & 0.100 &  31.0 & $-$24.5 &     &X&Y     \\
   8 & 12 &35.6 & 41 & 54 & 11 &15.97 &15.8 & KUV 08126+4154  & 1.280 &  32.9 & $-$28.7 &     & &Y     \\
   8 & 33 &34.0 & 44 & 36 & 30 &15.09 &15.5 & US 1329         & 0.249 &  37.0 & $-$25.6 &     &X&Y     \\
   9 & 46 &49.7 & 39 & 16 &  5 &15.97 &15.4 & KUV 09468+3916  & 0.360 &  50.7 & $-$25.8 &     &X&Y     \\
   9 & 53 &48.1 & 41 & 29 & 39 &15.59 &15.1 & PG 0953+415     & 0.239 &  51.7 & $-$25.1 &     &X&Y     \\
  10 &  7 &26.1 & 41 & 47 & 25 &15.97 &15.6 & FBS 1007+417    & 0.613 &  54.2 & $-$26.7 &     &X&Y     \\
  11 &  0 &27.4 & 77 & 15 &  8 &15.93 &15.6 & FBS 1100+774    & 0.313 &  38.6 & $-$25.4 &     &X&Y     \\
  11 &  2 &55.0 & 34 & 41 & 47 &16.00 &15.9 & FBS 1102+447    & 0.510 &  66.2 & $-$26.3 &     & &N     \\
  11 & 14 &20.0 & 44 & 29 & 57 &15.11 &14.8 & FBS 1114+444    & 0.144 &  64.5 & $-$24.4 &     & &Y     \\
  11 & 15 &46.0 & 40 & 42 & 19 &14.57 &14.8 & FBS 1115+407    & 0.154 &  66.7 & $-$25.1 &     &X&Y     \\
  11 & 21 &55.8 & 42 & 18 & 14 &15.84 &15.5 & FBS 1121+423    & 0.234 &  66.9 & $-$24.8 &     &X&Y     \\
  12 & 29 &28.3 & 71 &  0 & 47 &15.66 &15.3 & FBS 1229+710    & 0.208 &  46.3 & $-$24.7 &     &X&Y     \\
  12 & 29 &28.6 & 35 & 46 & 48 &15.24 &15.1 & CSO 900         & 0.131 &  80.6 & $-$24.1 & (4) &X&Y     \\
  13 &  3 &54.9 & 39 & 31 & 28 &15.71 &15.7 & FIRST J1306+3915& 0.447 &  77.5 & $-$25.8 &     & &Y     \\
  13 &  9 &58.4 & 35 & 31 & 15 &15.64 &15.7 & FBS 1309+355    & 0.183 &  80.7 & $-$24.4 & (4) &X&Y     \\
  13 & 12 &30.3 & 78 & 37 & 44 &15.84 &15.6 & HS 1312+7837    & 2.000 &  38.7 & $-$29.9 &     & &N     \\
  13 & 22 & 8.5 & 65 & 57 & 25 &15.71 &15.6 & FBS 1322+659    & 0.168 &  51.1 & $-$24.2 &     &X&Y     \\
  13 & 51 &15.6 & 36 & 35 & 33 &15.48 &15.3 & CSO 1022        & 0.286 &  73.9 & $-$25.6 &     & &Y     \\
  14 &  2 &37.7 & 43 & 41 & 27 &15.62 &15.1 & FBS 1402+436    & 0.320 &  68.0 & $-$25.7 &     & &Y     \\
  14 & 11 &50.0 & 44 & 14 & 12 &14.01 &13.9 & FBS 1411+442    & 0.089 &  66.4 & $-$24.5 &     &x&Y     \\
  14 & 44 &50.2 & 40 & 47 & 37 &15.45 &15.6 & FBS 1444+408    & 0.267 &  62.7 & $-$25.5 &     &X&Y     \\
  15 & 12 &46.8 & 37 &  1 & 55 &15.33 &15.9 & FBS 1512+370    & 0.369 &  58.3 & $-$26.3 &     &X&Y     \\
  16 & 21 &23.5 & 39 & 16 & 27 &15.91 &16.2 & B3 1621+392     & 1.970 &  44.7 & $-$29.8 &     &X&Y     \\
  16 & 24 &14.6 & 34 &  5 & 56 &15.80 &15.2 & RXS J16261+3359 & 0.204 &  43.8 & $-$24.5 &     &X&Y     \\
  16 & 34 &51.6 & 70 & 37 & 37 &15.27 &14.9 & FBS 1634+706    & 1.337 &  36.6 & $-$29.5 &     & &Y     \\
  16 & 41 &17.6 & 39 & 54 & 11 &15.88 &16.1 & FBS 1641+399    & 0.595 &  40.9 & $-$26.8 &     &X&Y     \\
  17 &  8 &23.2 & 33 & 47 & 42 &15.82 &15.6 & RXS J17102+3344 & 0.208 &  34.7 & $-$24.6 & (4) &X&Y     \\
  17 & 10 & 0.2 & 67 & 53 & 29 &15.94 &15.9 & HS 1710+6753    & 0.410 &  34.5 & $-$25.9 &     & &N     \\
  17 & 21 &32.0 & 34 & 20 & 41 &15.23 &15.3 & B2 1721+34      & 0.205 &  32.2 & $-$25.2 &     &X&Y     \\

\hline
\end{tabular}
\end{center}
\end{table*}
\normalsize

\noindent
In the case of RXS\,J12110$+$7005 for which the APS magnitude is not available, 
Schwope et al. (2000) give $B =$ 17.0, while the USNO $O$ magnitude is 14.3; but this 
object has a moderate redshift ($z =$ 0.127); moreover its APM $O$ magnitude (Irwin
et al. 1994) is 17.66; we therefore adopted the Schwope et al. mag and excluded
it from the ``complete" sample. 

\subsection{THE NEW RADIO AND X-RAY BRIGHT QSOs}

 The FIRST Bright Quasar Survey (FBQS) was built by matching the VLA FIRST
survey with the Cambridge Automated Plate Measuring Machine (APM) catalog of 
POSS-I objects (Irwin et al. 1994); it covers an area of 2\,682 deg$^{2}$ in the
north Galactic cap; it contains 1\,238 objects brighter than 17.8 mag on the 
POSS-I $E$ plates (White et al. 2000). About 1\,180 square degrees are within the
FBS area; they contain 38 FIRST radio sources identified with an AGN brighter 
than $B =$~17.0 at $|b| >$~30\degr; nine are bright QSOs 
($O_{\rm APS} <$~16.0), three (CSO~900, FIRST\,J1306$+$3915 and 
RXS\,J17102$+$3344) being new. 
Although the numbers are small, this suggests that the ``complete" sample we 
built in Paper I is only 67$\pm$15\/\% complete. 

 According to White et al. (2000), QSOs with radio emission above the FIRST 1 
mJy limit constitute about 25\/\% of all QSOs brighter than $B \sim$ 17.6, but 
for QSOs brighter than $B =$ 16.4, the FBQS QSO density is indistinguishable from 
the density of optically selected QSOs. Nevertheless, of the 15 bright QSOs 
known prior to the FIRST survey in the area common to the FIRST and FBS surveys,
only six (40\/\%) have been detected as FIRST radio sources; therefore the 
complete identification of the FIRST sources with bright starlike objects could 
not yield a complete survey of bright QSOs. \\

 A number of recent papers are devoted to the optical identification of RASS 
sources (Beuermann et al. 1999; Cao et al. 1999; Grazian et al. 2000; Schwope et
al. 2000; Wei et al. 1999; Xu et al. 1999). One of the new identifications is 
RXS\,J12043$+$4330, a QSO at $z =$~0.663 (Xu et al. 1999); it is also 
FBS\,1201$+$437 (FBS\,\#302) or PG\,1201$+$436, which had been classified as a 
DC white dwarf by Green et al. (1986). Its APS $O$ magnitude is 16.23; it is 
therefore not bright enough to be included in our ``complete" sample. \\

 Nineteen RASS sources are now identified with a bright QSO in the area
discussed in this paper (including the three new FIRST QSOs); of the 17 FBS or 
BQS bright QSOs in our sample (Table \ref{brightqso}), 12 (70\/\%) are ROSAT All 
Sky Survey (RASS) X-ray sources, suggesting that the total number of bright QSOs
is equal to 19/0.70 $=$ 27 (if all optically bright, X-ray sources have been 
discovered). 

\section{DISCUSSION}

 Our ``complete" sample of bright QSOs (Table \ref{brightqso}) contains 29 
objects brighter than $B =$ 16.16 ($O_{\rm APS} <$ 16.00), three of them (indicated 
by a ``N" in the last column of Table \ref{brightqso}) are not within the PG 
area. The area common to the PG and FBS surveys at $|b| >$ 30\degr\  
($\sim$2\,250 deg$^{2}$) contains 26 bright QSOs (13 PG QSOs and 13 others)
(but there are 17 PG QSOs with $B_{\rm PG} <$ 16.16 in the area; this larger 
number is probably due to the Eddington (1940) effect, the PG magnitudes being 
affected by relatively large errors, $\sigma \sim$ 0.37 mag). From these data,
we derive a surface density of 0.012 deg$^{-2}$, which is to be compared with 
the original value of the PG survey: 0.0064 deg$^{-2}$, implying a maximum 
completeness of 53$\pm$10\/\% for the PG survey. \\
 
 Grazian et al. (2000) have cross-correlated the RASS with photometric databases
in an 8\,164 deg$^{2}$ area of the northern sky at $|b|>$ 30\degr, 
selecting all coincidences brighter than $R \sim$~15.4; from this, they 
derive a surface density of bright ($B <$~15.5) QSOs (defined as AGNs with 
$M_{B}< -$23.0) of 10$\pm$2 10$^{-3}$ deg$^{-2}$ and conclude that the true 
surface density of such objects is about three times larger than that derived 
from the PG survey. However, they do not specify how the $B$ magnitude of their 
objects was derived. Their sample contains 46 QSOs; 15 of them have $z >$~0.20;
we have extracted from the APS catalogue the $O$ magnitudes for 12 of them (for 
the three others, these magnitudes are unavailable); it turns out that only one
(J172320.5$+$341756) has $O <$ 15.34, corresponding to $B <$~15.5, suggesting 
that the $O$ magnitudes used by Grazian et al. are underestimated and,
consequently, the surface density overestimated. \\

 Lamontagne et al. (2000) claim that they found a surface density of bright QSOs
three times larger than the PG value. They have searched for UV-excess 
stellar-like objects with $B <$ 16.5 and $U - B$ $< -$0.6 in a 840 deg$^{2}$ area 
covering the south Galactic cap; the errors in the $B$ magnitudes are estimated to
be 0.30 mag {\it rms}. They have found 228 such objects which have all been 
spectroscopically identified; 32 are AGNs, out of which only eleven are brighter
than $B =$~16.16 and $M_{B} = -$24.0 (including 0117$-$2837 which, according to 
Grupe et al. (1999), has a redshift of 0.349 rather than 0.055). We derive a 
surface density or 0.013 deg$^{-2}$, in agreement with our value and only twice 
the PG value.

\section{CONCLUSION}

 In Paper I, we compared the surface density of QSOs in the Bright Quasar Survey
and in the First Byurakan Survey and concluded that the completeness of the BQS 
is of the order of 70\/\%; Wisotzki et al. (2000) have found that the BQS is 68\/\%
complete from a comparison with the Hamburg/ESO survey, in agreement with our 
previoys estimate.
 Based on a number of recently published data, as well as on our own new 
observations, we redetermined the surface density of QSOs brighter than $B =$~16.16
in the BQS area to be $\sim$0.012 deg$^{-2}$, implying that the completeness
of the BQS is 53$\pm$10\/\%. It should be stressed however that the numbers 
involved are quite small, and that larger areas should be investigated before a 
definitive value of the surface density of bright QSOs could be determined.


\end{document}